\documentclass[draft]{agujournal2019}
\usepackage{url} 
\usepackage{lineno}
\usepackage[inline]{trackchanges} 
\usepackage{soul}
\usepackage{siunitx}
\DeclareSIUnit\year{year}
\DeclareSIUnit\years{years}
\DeclareSIUnit\ppm{ppm}
\draftfalse

\journalname{JGR: Machine Learning and Computation}

\begin{document}

%
%


\title{ACE2-SOM: Coupling an ML atmospheric emulator to a slab ocean and learning the sensitivity of climate to changed CO$_2$}

%
%




\authors{Spencer K. Clark\affil{1,2}, Oliver Watt-Meyer\affil{1}, Anna Kwa\affil{1}, Jeremy McGibbon\affil{1}, Brian Henn\affil{1}, W. Andre Perkins\affil{1}, Elynn Wu\affil{1}, Lucas M. Harris\affil{2}, and Christopher S. Bretherton\affil{1}}


\affiliation{1}{Allen Institute for Artificial Intelligence, Seattle, WA, USA}
\affiliation{2}{NOAA/Geophysical Fluid Dynamics Laboratory, Princeton, NJ, USA}




\correspondingauthor{Spencer K. Clark}{spencerc@allenai.org}



\begin{keypoints}
\item The Ai2 Climate Emulator coupled to a slab ocean accurately emulates temperature and precipitation CO$_2$ sensitivity in a physics-based model
\item Inference in an out-of-sample scenario with gradually increasing CO$_2$ is also accurate except for regime shifts in its stratosphere 
\item Abrupt 4xCO$_2$ inference reaches the correct equilibrium climate but the atmosphere warms too fast due to energy non-conservation
\end{keypoints}

%
%

%
%


\begin{abstract}

While autoregressive machine-learning-based emulators have been trained to produce stable and accurate rollouts in the climate of the present-day and recent past, none so far have been trained to emulate the sensitivity of climate to substantial changes in CO$_2$ or other greenhouse gases.  As an initial step we couple the Ai2 Climate Emulator version 2 to a slab ocean model (hereafter ACE2-SOM) and train it on output from a collection of equilibrium-climate physics-based reference simulations with varying levels of CO$_2$.  We test it in equilibrium and non-equilibrium climate scenarios with CO$_2$ concentrations seen and unseen in training.

ACE2-SOM performs well in equilibrium-climate inference with both in-sample and out-of-sample CO$_2$ concentrations, accurately reproducing the emergent time-mean spatial patterns of surface temperature and precipitation change with CO$_2$ doubling, tripling, or quadrupling. In addition, the vertical profile of atmospheric warming and change in extreme precipitation rates up to the 99.9999th percentile closely agree with the reference model. Non-equilibrium-climate inference is more challenging.  With CO$_2$ increasing gradually at a rate of \qty{2}{\percent \per \year}, ACE2-SOM can accurately emulate the global annual mean trends of surface and lower-to-middle atmosphere fields but produces unphysical jumps in stratospheric fields.  With an abrupt quadrupling of CO$_2$, ML-controlled fields transition unrealistically quickly to the 4xCO$_2$ regime. In doing so they violate global energy conservation and exhibit unphysical sensitivities of and surface and top of atmosphere radiative fluxes to instantaneous changes in CO$_2$.  Future emulator development needed to address these issues should improve its generalizability to diverse climate change scenarios.

\end{abstract}

\section*{Plain Language Summary}

Machine-learning-based models of the atmosphere have proven to be many times faster than traditional state of the art numerical weather prediction models and in many cases more accurate.  In the last few years there has been progress toward using similar approaches to accelerate climate simulations.  However, none so far have taken on the challenge of simulating the climate response to substantial increases in carbon dioxide.  In this study we build upon the latest version of our climate emulator, the Ai2 Climate Emulator version 2, to work toward addressing this problem.  We connect our emulator to a simplified physics-based model of the ocean, and train on output from a set of physics-based climate model simulations with one, two, and four times the present-day carbon dioxide concentration.  With this approach, our new model, which we call ACE2-SOM, emulates equilibrium changes in temperature and precipitation as well or better than a 25x-less energy-efficient physics-based model.  It struggles, however, in emulating the rate of climate adjustment to a new carbon dioxide concentration, generally doing so too quickly.  We speculate that addressing this will require selectively building more physics into the model, but we believe that is a good opportunity for future work.

%
%

\section{Introduction}

A number of studies have demonstrated autoregressive machine-learning-based emulators of global atmosphere models can produce stable and accurate multi-year simulations of climate with a tiny fraction of the time and computer resources required by traditional physics-based models \cite{Wey2020,Wat2023a,Dun2024,Wat2024,Gua2024,Kar2024,Cre2024}. A limitation of these emulators is that they have all been trained on either ERA5 reanalysis data \cite{Wey2020,Gua2024,Kar2024,Cre2024,Wat2024} or physics-based model output with present-day annually repeating \cite{Wat2023a,Dun2024} or annually varying observed sea surface temperature (SST) and sea ice boundary conditions \cite{Wat2024}.  Limitations of fully data-driven or hybrid models in making reliable predictions of data outside the range seen during training \cite<e.g.,>{OG2018,Koc2024,Lin2024,Rac2024} restrict these models' applicability to emulating the climate of roughly the last 80 years. This has utility to seasonal forecasting, but a critical application of climate models is to simulate what climate would be like under anthropogenic forcings outside the historical range.

While climate is known to be sensitive to a variety of different forcing agents, e.g. different types of greenhouse gases, aerosols, or land use changes \cite{ON2016,Ria2017}, we simplify this study by aiming to emulate the response of climate to changes in CO$_2$ alone. There is a long history of physics-based climate modeling experiments in this vein. For example, abrupt and gradually increasing CO$_2$ experiments have been a central part of the past three Coupled Model Intercomparison Projects \cite{Mee2007,Tay2012,Eyr2016}. Such experiments, along with equilibrium climate simulations with perturbed CO$_2$, can serve as a useful framework for assessing and discussing how well our emulator captures the physical response of different aspects of climate to an important greenhouse gas. These aspects include (but are not limited to) the time-mean spatial pattern of the change in temperature and precipitation between climates with present-day and perturbed CO$_2$, the impact of CO$_2$ on surface and top of atmosphere radiative fluxes, and the response of climate to a time-evolving CO$_2$ concentration.

Our starting point is the Ai2 Climate Emulator version 2 (ACE2) as described in detail in \citeA{Wat2024}. To produce reference output for training, validation, and testing we run GFDL's SHiELD model \cite{Har2020} with a variety of different CO$_2$ concentrations. Since altering the CO$_2$ concentration will naturally be expected to change the sea surface temperature (SST), we couple SHiELD to a slab ocean. This is a simplified physics-based ocean model in which heat fluxes due to ocean circulation are prescribed, but heat exchange with the atmosphere is interactive, allowing the SST to respond to changes in the radiative and turbulent energy fluxes at the surface caused directly or indirectly by CO$_2$. Coupling to such a simple ocean has the virtue that it is straightforward to implement a differentiable version in ACE2 that can be used during training and inference. A slab ocean also equilibrates orders of magnitude more quickly than a dynamical ocean, making it more efficient for generating reference data in multiple climates \cite<e.g.,>{Dan2009}. In this project we therefore train ACE2 coupled to a slab ocean (ACE2-SOM) to emulate SHiELD coupled to a slab ocean (SHiELD-SOM) with varying levels of CO$_2$, leaving more advanced treatment of ocean dynamics to future work.

There is an extensive literature about approaches to infer how physics-based climate models would respond to different emissions scenarios of greenhouse gases and aerosols based on a set of reference runs.  These range from simple approaches like pattern scaling \cite{San1990,Mit2003} to more advanced approaches like deep learning models \cite{Wat2022,Ngu2023}, or impulse response functions \cite{Fre2024,Wom2024} to emulate a variable's annual mean time series. Approaches also exist to temporally downscale these predictions to monthly or finer resolution \cite{Nat2022,Bas2024}. For certain applications, like predicting the mean spatial pattern of surface temperature change under the Shared Socioeconomic Pathway 245 emissions scenario \cite{ON2016,Ria2017}, these can be quite accurate \cite{Wat2022,Ngu2023,Lut2024,Wom2024}.  However, they have restrictions—they emulate statistics rather than weather directly, often focus on a small set of variables \cite{Sch2024}, and they may not generalize across climate change scenarios.

ACE is different in that it is designed much like a climate model itself.  It autoregressively predicts the evolution of a coherent suite of meteorological variables with a short timestep, allowing for the explicit simulation and characterization of emergent weather and climate phenomena.  It is more computationally expensive and requires more training data, but with future development it could potentially become a complementary option for those interested in an emulator that can produce a richer, more interpretable, set of climate statistics.  In this pilot study we focus on evaluating the performance of ACE2-SOM as a computationally inexpensive climate model, using a coarse-resolution comprehensive physics-based model as a baseline, rather than compare to statistical emulator approaches.

We begin by describing the reference simulations we run with SHiELD, as well as the details of how we configure and train ACE2-SOM in Section~\ref{sec:data-and-methods}.  We then describe inference results in both equilibrium and non-equilibrium climates in Section~\ref{sec:results}. Our primary focus is on cases where the CO$_2$ concentration was not seen during training.  We highlight both aspects of emulation that ACE2-SOM does well, as well as opportunities for improvement in future work that we finally expand upon in Section~\ref{sec:discussion-and-conclusion}.

\section{Data and Methods}
\label{sec:data-and-methods}

\subsection{Reference physics-based simulations}

To generate reference data in multiple climates for training, validation, and testing, we make use of GFDL's SHiELD model \cite{Har2020}. We start from the same base configuration used for the AMIP reference simulations described in \citeA{Wat2024}, a 79-vertical-level model with physical parameters configured following what was used in the X-SHiELD simulations in \citeA{Che2022}, with the latest versions of both the shallow and deep convection schemes active. Using the latest versions of the convection schemes is important because the prior deep convection scheme was prone to instability in climates with increased CO$_2$. Unlike \citeA{Wat2024}, our reference runs include a slab ocean model instead of using prescribed SSTs.

\subsubsection{Slab ocean model}
\label{sec:som}

A slab ocean model (SOM) is a simplified physics-based model of the ocean. It approximates the near-surface ocean as a single well-mixed layer of water with a prescribed spatiotemporally-varying depth, whose temperature evolves through energy exchange with the atmosphere and the prescribed spatiotemporally-varying effect of dynamical ocean heat transport. The equation governing the ocean mixed layer temperature implemented in SHiELD follows \citeA{Kie2006}, which has its roots in a model used by \citeA{Han1983}:
\begin{equation}
    \label{eq:som}
    \rho_o C_o h \frac{\partial T_s}{\partial t} = F_{net} + Q.
\end{equation}
Here $\rho_o = \qty{1000}{\kg \per \m \cubed}$ and $C_o = \qty{4000}{\J \per \kg \per \K}$ are the density and specific heat of water, respectively; $h$ is the prescribed mixed layer depth; $T_s$ is the mixed layer temperature, which by way of mixing is equivalent to the ocean surface temperature; $F_{net}$ is the net downward surface energy flux; and $Q$ is the prescribed flux convergence due to ocean heat transport, hereafter the ``Q-flux.''  $F_{net}$ can be expressed in terms of the radiative and turbulent fluxes at the surface:
\begin{equation}
    F_{net} = R^{lw}_{down} - R^{lw}_{up} + R^{sw}_{down} - R^{sw}_{up} - SH - LH,
\end{equation}
where $R^{lw}_{down}$ and $R^{lw}_{up}$ are the downward and upward components of the longwave radiative flux, $R^{sw}_{down}$ and $R^{sw}_{up}$ are the downward and upward components of the shortwave radiative flux, $SH$ is the sensible heat flux, and $LH$ is the latent heat flux.

We use a prescribed mixed layer depth climatology produced by \citeA{de2004}, who inferred it from multiple observational sources from 1941 through 2002. It is a monthly mean climatology on a $\qty{2}{\degree} \times \qty{2}{\degree}$ regular latitude-longitude grid. The Q-flux climatology is derived from this mixed layer depth climatology, combined with 30 years of output from SHiELD run with prescribed annually repeating climatological monthly mean SSTs \cite{Thi2003} and sea ice \cite{Sah2014} for the period 1982 to 2012.  In these simulations CO$_2$ is set perpetually to the observed concentration in year 1997, \qty{363.43}{\ppm}, which we refer to hereafter as ``1xCO$_2$.'' To derive an implied climatological Q-flux we solve Equation~\ref{eq:som} for $Q$ using the mixed layer depth climatology, the climatological monthly mean prescribed SST, and the simulated $F_{net}$. To align the mixed layer depth climatology with the grid of the prescribed SST and simulated $F_{net}$, we first fill missing values with nearest neighbor interpolation on the sphere and then regrid using bilinear interpolation, leveraging ideas and code from the WeatherBench 2 project \cite{Ras2023}. This Q-flux derivation ensures that the climatological mean SST in an otherwise identically configured SHiELD simulation coupled to a slab ocean will approximately match that of the prescribed SST climatology in the reference case.

\subsubsection{Treatment of sea ice}
\label{sec:sea-ice}

While some atmosphere models coupled to a slab ocean include simplified interactive models of sea ice \cite<e.g.,>{Kie2006}, we prescribe sea ice based on the same annually repeating observational climatology used in the prescribed SST reference simulations for computing the Q-flux \cite{Sah2014}. We acknowledge that running with prescribed sea ice, while simplifying our setup, eliminates an important amplifying climate change feedback mechanism, particularly in the high latitudes.  For example, \citeA{Hal2004} showed that ignoring the ice-albedo feedback for both ocean and land reduced the surface air temperature increase in response to a doubling of CO$_2$ in high-latitude regions by a factor between about \num{1.4} and \num{2.2}.  We later use the same sea-ice treatment in our emulator as we do in SHiELD for consistency.

\subsubsection{Simulation protocol}

The full suite of SHiELD simulations completed for this study is summarized in Table~\ref{tab:shield-simulations}. We run all of these simulations at both C96 (roughly \qty{100}{\km}) and C24 (roughly \qty{400}{\km}) resolution. The C96 simulations produce the target output we seek to emulate, while the C24 simulations serve as a computationally inexpensive physics-based baseline for comparison.  We tune down the strength of the mountain blocking scheme in the C24 resolution simulations as in \citeA{Wat2024} to reduce their climate biases relative to those at C96 resolution, based on the scheme's previously documented sensitivity to resolution (J. Alpert and F. Yang, personal communication, August 9, 2019). For equilibrium-climate simulations with annually repeating forcings we use an ensemble approach to parallelize data generation; in Table~\ref{tab:shield-simulations} these are groups of simulations with an ensemble size greater than \num{1}. To generate unique ensemble members, we use a similar strategy to \citeA{Des2012}: force-starting the model with different initial conditions selected from the final days of the relevant spin-up run, as noted in the initial condition column. We ignore some spin-up time in ensemble simulations to allow the different members to diverge; the spin-up period in runs with prescribed SSTs is \num{3}~months, while the spin-up period in slab ocean runs is \num{1}~year.

The workflow at each resolution starts with running prescribed SST simulations to produce a reference Q-flux climatology. For this purpose we use \num{30} post-spin-up years spread across two ensemble members. The \qty{15}{\year} spin-up period prior is mainly required for the stratospheric water vapor to equilibrate after being initialized from GFS analysis. With a Q-flux climatology computed following the approach described in Section~\ref{sec:som}, we then run spin-up slab ocean simulations. This begins with spinning up the slab ocean in the 1xCO$_2$ climate with a \qty{10}{\year} run initialized off of the end of one of the prescribed SST simulation ensemble members. From the end of that run, we then initialize three \qty{10}{\year} abrupt CO$_2$ change simulations with 2x, 3x, and 4xCO$_2$. These simultaneously serve as spin-up simulations for the model in each of these climates, as well as reference cases with abrupt CO$_2$ change. Finally we initialize five-member ensembles of 10~post-spin-up-year equilibrium climate runs off the ends of the spin-up simulations in each climate, as well as initialize a single 70~post-spin-up-year simulation with CO$_2$ starting at the 1xCO$_2$ level and increasing at a rate of \qty{2}{\percent \per \year} to the 4xCO$_2$ level.

In summary, our reference dataset produced with SHiELD-SOM at each horizontal resolution consists of \num{50}~years of equilibrium climate simulation output with each of 1xCO$_2$, 2xCO$_2$, 3xCO$_2$, and 4xCO$_2$, \num{10}~years of abrupt CO$_2$ change from 1xCO$_2$ simulation output with each of a CO$_2$ doubling, tripling, and quadrupling, and finally \num{70}~years of gradual CO$_2$ change simulation output with CO$_2$ increasing at a rate of \qty{2}{\percent \per \year}.

\begin{sidewaystable}
\caption{Summary of SHiELD simulations completed for this study.}
\centering
\begin{tabular}{c c c c c c c}
\hline
 Name  & Ocean & CO$_2$ & Resolutions & Initial condition$^{a}$ & Time span$^{b}$ & Ensemble size$^{c}$ \\
\hline
climSST spin-up & Data & 1xCO$_2$ & C96, C24 & GFS analysis for 2020-01-01 00Z & 2020-01-01 to 2034-10-01 & 1 \\
Q-flux reference & Data & 1xCO$_2$ & C96, C24 & End of climSST spin-up & 2034-10-01 to 2050-01-01 & 2 \\
\hline
1xCO$_2$ spin-up & Slab & 1xCO$_2$ & C96, C24 & End of Q-flux reference & 2020-01-01 to 2030-01-01 & 1 \\
Abrupt 2xCO$_2$ & Slab & 2xCO$_2$ & C96, C24 & End of 1xCO$_2$ spin-up & 2020-01-01 to 2030-01-01 & 1 \\
Abrupt 3xCO$_2$ & Slab & 3xCO$_2$ & C96, C24 & End of 1xCO$_2$ spin-up & 2020-01-01 to 2030-01-01 & 1 \\
Abrupt 4xCO$_2$ & Slab & 4xCO$_2$ & C96, C24 & End of 1xCO$_2$ spin-up & 2020-01-01 to 2030-01-01 & 1 \\
\hline
1xCO$_2$ & Slab & 1xCO$_2$ & C96, C24 & End of 1xCO$_2$ spin-up & 2030-01-01 to 2041-01-01 & 5 \\
2xCO$_2$ & Slab & 2xCO$_2$ & C96, C24 & End of Abrupt 2xCO$_2$ & 2030-01-01 to 2041-01-01 & 5 \\
3xCO$_2$ & Slab & 3xCO$_2$ & C96, C24 & End of Abrupt 3xCO$_2$ & 2030-01-01 to 2041-01-01 & 5 \\
4xCO$_2$ & Slab & 4xCO$_2$ & C96, C24 & End of Abrupt 4xCO$_2$ & 2030-01-01 to 2041-01-01 & 5 \\
\hline
Increasing CO$_2$ & Slab & \qty{2}{\percent \per \year} increase$^{d}$ & C96, C24 & End of 1xCO$_2$ spin-up & 2030-01-01 to 2101-01-01 & 1 \\
\hline
\multicolumn{7}{p{22cm}}{$^{a}$Unique initial conditions are used for each ensemble simulation. Multiple initial conditions from the same source are derived by saving daily restart files during the last month of the source simulation.}\\
\multicolumn{7}{p{22cm}}{$^{b}$Where the start time and initial condition time differ, the initial condition is treated as though it occurs at the start time. This is important in particular for generating unique ensemble members in the case that the configuration of the ensemble runs is identical to that used to generate the initial conditions.}\\
\multicolumn{7}{p{22cm}}{$^{c}$For ensemble sizes greater than one, \num{3}~months of spin-up time from each run are ignored to allow states to diverge when running with a data ocean, and \num{1}~year of spin-up time from each run is ignored when running with a slab ocean.}\\
\multicolumn{7}{p{22cm}}{$^{d}$The first two years are run with 1xCO$_2$, with the first year meant to allow the state to diverge from the spin-up simulation. CO$_2$ increases at a rate of \qty{2}{\percent \per \year} thereafter up to about 4xCO$_2$ in 2100.}
\end{tabular}
\label{tab:shield-simulations}
\end{sidewaystable}

\subsubsection{Data pre-processing}

To prepare the data output from these simulations for use with ACE2, we follow the same pre-processing procedure described in \citeA{Wat2024}. Using \texttt{fregrid} \cite{NOA2024}, C96 and C24 output are regridded to \qty{1}{\degree} and \qty{4}{\degree} Gaussian grids respectively, and then all but the surface type fraction variables are run through a spherical harmonic transform (SHT) round trip. Finally vertically resolved fields (air temperature, specific total water, eastward wind, and northward wind) are conservatively remapped via mass-weighted averages from SHiELD's native \num{79} hybrid vertical levels to ACE2's \num{8} hybrid levels.
 
\subsection{Implementation, training, and testing of ACE2-SOM}

The output from the physics-based reference simulations is used for training and testing ACE2 coupled to a slab ocean model (ACE2-SOM). Other than the slab ocean, we configure ACE2-SOM identically to ACE2 as described in \citeA{Wat2024}. It uses the same ML architecture \cite<the Spherical Fourier Neural Operator architecture introduced in>{Bon2023}, grid (\qty{1}{\degree} horizontal resolution with \num{8} vertical layers), embedding dimension size (\num{384}), ML input and output variables (see Table~S1 for a description of those in the context of this study), variable normalization approach, and loss function, and it enforces exact conservation of global dry air and column moisture within the atmosphere. We refer the reader there for a more detailed description of each of those aspects.

\subsubsection{Slab ocean implementation}

We implement coupling to a slab ocean model as a configuration option in ACE2. Like the rest of ACE2, the slab ocean component is written within the \texttt{PyTorch} framework, so the full model remains differentiable for optimal training. The main difference between ACE2 and ACE2-SOM is that for each \num{6}-hour prediction timestep, the predicted \num{6}-hour mean surface fluxes that comprise $F_{net}$, the mixed layer temperature at the start of the timestep, and the prescribed Q-flux and mixed layer depth for the timestep, are supplied to the slab ocean model to update the mixed layer (and ocean surface) temperature at the end of the timestep.  

Naturally the slab ocean model is only applicable for use in ocean grid cells. We handle this in ACE2-SOM by allowing both the slab ocean model and the ML to predict the surface temperature globally. The surface temperature produced by the coupled model at the end of each timestep is computed as the weighted average of the two based on the fraction of the area of the grid cell covered by ocean, $f_o$:
\begin{equation}
    T_s = f_o T^{SOM}_s + (1 - f_o) T^{ML}_s.
\end{equation}
This is the surface temperature fed back as an input in the next timestep, and that used in computing the training loss. Unlike in SHiELD, grid cells with fractional ocean are possible in ACE2-SOM due to regridding from SHiELD's native cubed sphere grid to ACE2-SOM's Gaussian grid.

\subsubsection{Training}

We train and validate on data from the 1xCO$_2$, 2xCO$_2$, and 4xCO$_2$ equilibrium climates, leaving output from the remaining reference simulations for out-of-sample testing. We select \num{40} years of data—the first four ensemble members—from each of the 1xCO$_2$, 2xCO$_2$, and 4xCO$_2$ climates for training, leaving the remaining \num{10} years in each climate for validation, and compute normalization statistics using the training dataset. We train ACE2-SOM for \num{30} epochs, since we find that while training loss may not yet converge by epoch \num{30}, inference skill peaks typically before epoch \num{15} with this dataset. As in \citeA{Wat2024}, we run a suite of inference simulations at the end of each epoch to help select a checkpoint based on best climate skill. In our case, we run five-year simulations with \num{8} different initial conditions selected from each of the validation datasets for the 1xCO$_2$, 2xCO$_2$, and 4xCO$_2$ climates, i.e. \num{24} inference runs per epoch. Consistent with \citeA{Wat2024}, we found that selecting a checkpoint based on climate skill with a range of in-sample forcings was most predictive of climate skill with unseen forcings. We train models with four different random seeds, and focus on results with the model that produced the best inline inference skill.

\subsubsection{Testing}

We test ACE2-SOM by running an analogous suite of simulations to that performed with SHiELD-SOM (Table~\ref{tab:shield-simulations}). Five-member ensembles of equilibrium-climate runs are generated by initializing ACE2-SOM with conditions selected from the first five timesteps at the start of the spin-up period of a reference ensemble member in each of the 1xCO$_2$, 2xCO$_2$, 3xCO$_2$, and 4xCO$_2$ climates. The simulations are given a year to diverge from each other and run for \num{10} years thereafter, which we use for analysis. We also run ACE2-SOM simulations with CO$_2$ increasing at a rate of \qty{2}{\percent \per \year}, and CO$_2$ concentration abruptly quadrupled from 1xCO$_2$ to 4xCO$_2$.

The 3xCO$_2$ equilibrium climate, CO$_2$ quadrupling, and gradual CO$_2$ increase runs are all forced with CO$_2$ concentrations and/or combinations of CO$_2$ concentrations and atmospheric states that were not seen during training; we refer to these as out-of-sample test cases. The 1xCO$_2$, 2xCO$_2$, and 4xCO$_2$ simulations can be considered in-sample test cases at least from the perspective of the CO$_2$ concentration and character of the corresponding atmospheric states. In the following discussion we focus mainly on results from the more challenging out-of-sample test cases, though we will touch briefly on results from in-sample cases.

\subsection{Computational cost}
Since we use the same hardware and processor layouts, and running SHiELD or ACE2 with a slab ocean costs roughly the same as running with prescribed SSTs, the computational cost and energy use rate of our simulations is similar to that reported in \citeA{Wat2024}. C96 SHiELD-SOM simulation throughput is therefore roughly \num{11.4}~simulated years per day, with an energy use rate of \qty{8250}{Wh} per simulated year; C24 SHiELD-SOM simulation throughput is roughly \num{22.1}~simulated years per day, with an energy use rate of \qty{300}{Wh} per simulated year; and ACE2-SOM runs at a rate of roughly \num{1500}~simulated years per day, with an energy use rate of \qty{11.2}{Wh} per simulated year, approximately \num{100} times faster and \num{700} times more energy efficient than its target model. We train ACE2-SOM on 8 NVIDIA H100 GPUs; each epoch takes about \qty{6000}{seconds}, meaning training for \num{30} epochs takes about \qty{50}{hours} and uses \qty{280000}{Wh} of electricity.

\section{Results}
\label{sec:results}

\subsection{Equilibrium climate inference}

\subsubsection{Skill in emulating individual climates}

\begin{figure}
\noindent\includegraphics[width=\textwidth]{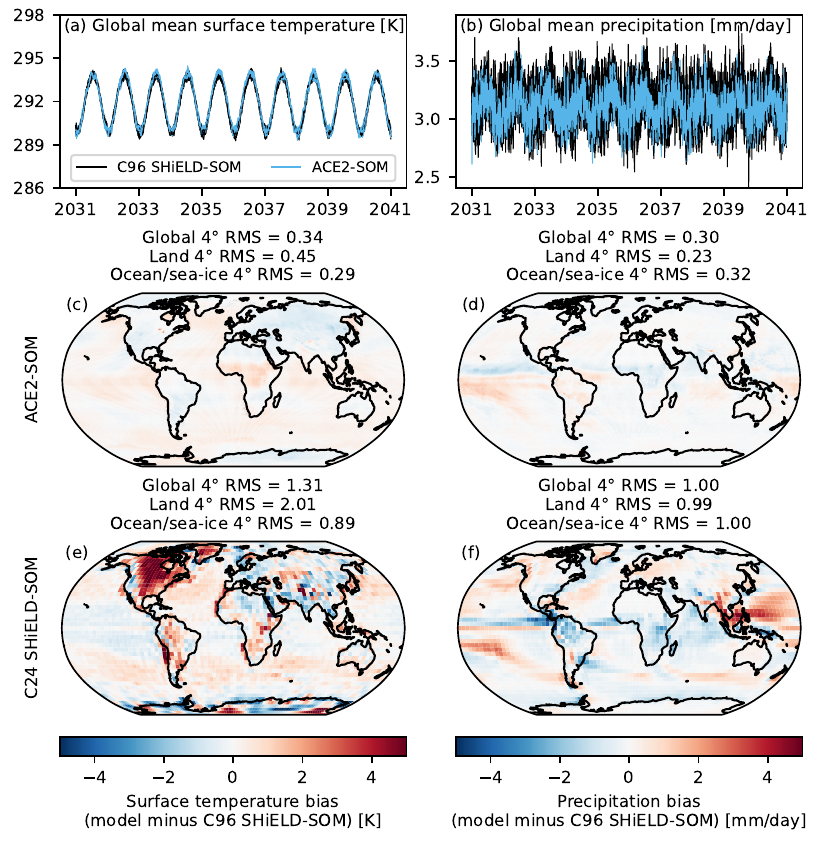}
\caption{Time series of daily and global mean surface temperature (a) and precipitation (b) with 3xCO$_2$ in each ensemble member of C96 SHiELD-SOM (black) and ACE2-SOM (blue). Time and ensemble mean bias in surface temperature (c) and precipitation (d) in ACE2-SOM relative to C96 SHiELD-SOM, and the same for C24 SHiELD-SOM relative to C96 SHiELD-SOM in (e) and (f). Note that ACE2-SOM bias maps are plotted at \qty{1}{\degree} resolution, while C24 SHiELD-SOM bias maps are plotted at \qty{4}{\degree} resolution. Root mean square (RMS) metrics, however, are always reported at \qty{4}{\degree} resolution to ensure a fair comparison between the two.}
\label{fig:general-3xCO2-climate}
\end{figure}

To illustrate the stability and accuracy of ACE2-SOM, we first plot the time series of daily and global mean surface temperature and precipitation from each ensemble member in the out-of-sample 3xCO$_2$ climate compared with that of C96 SHiELD-SOM in Figures~\ref{fig:general-3xCO2-climate}a~and~\ref{fig:general-3xCO2-climate}b. It is evident that all ensemble members of ACE2-SOM follow the global mean annual cycle of the unseen target well, and none exhibit unusual deviations, systematic global mean biases, or temporal drift. Time and ensemble mean bias maps of ACE2-SOM relative to C96 SHiELD-SOM, shown in Figures~\ref{fig:general-3xCO2-climate}c~and~\ref{fig:general-3xCO2-climate}d for surface temperature and precipitation, also indicate that ACE2-SOM's biases are small in all regions. In contrast, Figures~\ref{fig:general-3xCO2-climate}e~and~\ref{fig:general-3xCO2-climate}f show that our physics-based baseline, C24 SHiELD-SOM, has much larger biases in land surface temperature and tropical precipitation.  After conservative regridding to a common \qty{4}{\degree} Gaussian grid, ACE2-SOM reduces global root mean squared error in time-mean surface temperature by \qty{74}{\percent} and in time-mean precipitation by \qty{70}{\percent} vs. the baseline.

\begin{figure}
\noindent\includegraphics[width=\textwidth]{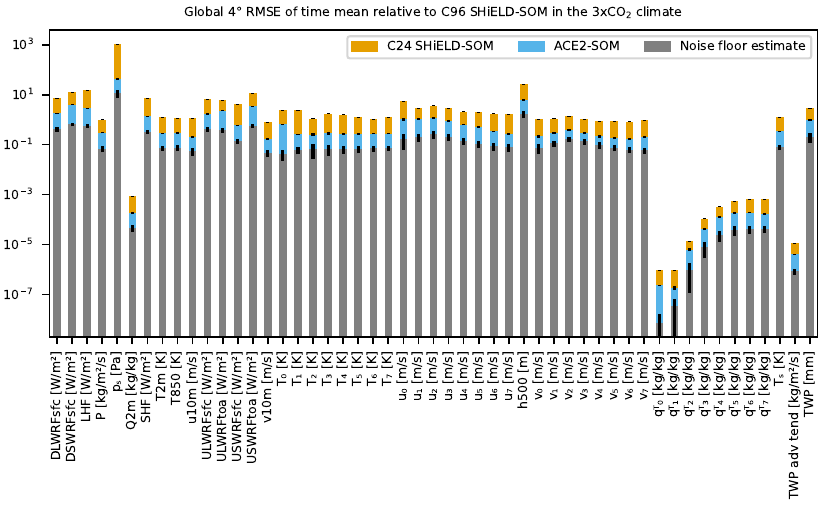}
\caption{\qty{4}{\degree} root mean square error of the time and ensemble mean for all variables predicted by ACE2-SOM (blue), compared to that for C24 SHiELD-SOM (orange) and a noise floor estimate (gray). Error bars represent $\pm \, 2$ standard deviations of the noise floor. The uncertainty is assumed to be similar for ACE2-SOM and C24 SHiELD-SOM, so we use the same error bars, though the logarithmic scale of the $y$-axis makes the size of the error bars appear different. See Table~S1 for a summary of the variables predicted by ACE2-SOM and their associated short names.}
\label{fig:rmse-bar-chart-3xCO2}
\end{figure}

This impressive climate emulation skill holds for all fields predicted by ACE2-SOM. Figure~\ref{fig:rmse-bar-chart-3xCO2} shows that the global \qty{4}{\degree} root mean square error of the time and ensemble mean in the 3xCO$_2$ climate for all diagnostic and prognostic ACE2-SOM fields is smaller than that in the C24 SHiELD-SOM baseline by between \qtyrange[range-units = single]{54}{96}{\percent} depending on the variable. ACE2-SOM is not a perfect emulator, however, indicated by the fact that its RMSE of the time and ensemble mean for all fields is still larger than that of a ``noise floor'' estimate of the error statistics that another independent 50-year ensemble of C96 SHiELD-SOM simulations might produce. The noise floor estimate is calculated by computing the global RMSE of the time mean for \num{5} and \num{10} year windows of reference data relative to the full \num{50} years available, and fitting a curve of the form:
\begin{equation}
    RMSE_v(N) = \frac{RMSE_v(1)}{\sqrt{N}}
\end{equation}
to extrapolate the value of the RMSE for each variable $v$ for $N = \qty{50}{\years}$. Error bars, representing a roughly \qty{95}{\percent} confidence interval, are $\pm \, 2$ standard deviations of the RMSE across windows, extrapolated to the 50-year case in a similar way. The noise floor accounts for the fact that simulated climates with the same model have significant internal variability, which is expected to be uncorrelated between separate multi-year periods. The picture is similar if we look at single-climate skill in the in-sample 1xCO$_2$, 2xCO$_2$, and 4xCO$_2$ climates, where ACE2-SOM's \qty{4}{\degree} RMSE improvements over the baseline are \qtyrange[range-units = single]{56}{94}{\percent}, \qtyrange[range-units = single]{75}{97}{\percent}, and \qtyrange[range-units = single]{67}{97}{\percent}, respectively.

\subsubsection{Skill in emulating climate change patterns}

\begin{figure}
\noindent\includegraphics[width=\textwidth]{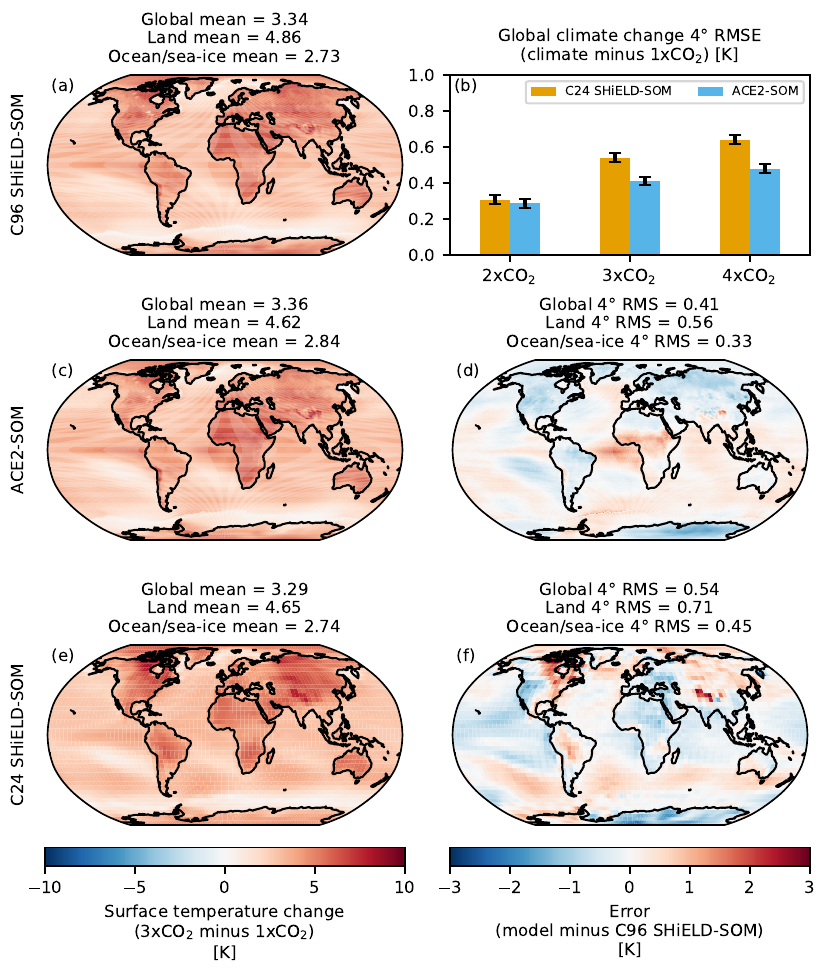}
\caption{Time and ensemble mean difference in surface temperature between the 3xCO$_2$ climate and 1xCO$_2$ climate in C96 SHiELD-SOM (a), ACE2-SOM (c), and C24 SHiELD-SOM (e). Panels (d) and (f) show the error in emulating this change pattern for ACE2-SOM and C24 SHiELD-SOM, respectively. Panel (b) shows the global \qty{4}{\degree} RMSE of the climate change pattern for all the perturbed climates for ACE2-SOM and C24 SHiELD-SOM relative that for C96 SHiELD-SOM. Error bars represent $\pm$ \qty{2} standard deviations of the RMSE of the difference between climates; here the standard deviation of the RMSE of the difference between climates is computed as $\sqrt{2}$ times the standard deviation of the RMSE in a single climate.}
\label{fig:surface-temperature-change-3xCO2}
\end{figure}

Since ACE2-SOM emulates the time and ensemble mean pattern of variables well in each individual climate, it also accurately emulates climate change patterns. The left column of Figure~\ref{fig:surface-temperature-change-3xCO2} shows the difference in time and ensemble mean surface temperature between the 3xCO$_2$ and 1xCO$_2$ climate in C96 SHiELD-SOM, ACE2-SOM, and C24 SHiELD-SOM simulations. All exhibit a qualitatively similar pattern. There is an El Ni\~no-like warming pattern in the tropical east Pacific Ocean, which is a common feature of many physics-based models \cite<e.g.,>{Son2014,Kan2023}, though its consistency with observations and physical mechanism is still a topic of research \cite{Lee2022}. Land warms more relative to ocean/sea-ice, which is a ubiquitous feature of physics-based models with a well-understood physical mechanism \cite<e.g.,>{Sut2007}. Finally, there is little to no warming over regions of sea ice, which as discussed in Section~\ref{sec:sea-ice} is prescribed, and therefore held fixed, unlike in typical comprehensive coupled climate models where it can feed back with changes in climate \cite<e.g.,>{Hel2019,Gol2022}.

Figures~\ref{fig:surface-temperature-change-3xCO2}d~and~\ref{fig:surface-temperature-change-3xCO2}f show maps of the error in emulating the climate change pattern of surface temperature in the 3xCO$_2$ climate relative to C96 SHiELD-SOM for ACE2-SOM and C24 SHiELD-SOM. Both ACE2-SOM and C24 SHiELD-SOM emulate this change pattern well, with global \qty{4}{\degree} RMSEs of less than \qty{0.6}{\kelvin}. While C24 SHiELD-SOM exhibits large biases in individual climates, these biases are relatively consistent, so when taking the difference between climates they largely cancel out.  Nevertheless, the pattern RMSE of the temperature change bias is smaller for ACE2-SOM than for the C24 SHiELD-SOM baseline. Figure~\ref{fig:surface-temperature-change-3xCO2}b compares the global climate change pattern RMSEs between ACE2-SOM and C24 SHiELD-SOM, for each of the in-sample and out-of-sample perturbed climates. It shows that ACE2-SOM robustly emulates the time-and-ensemble mean climate change pattern of C96 SHiELD-SOM more closely than C24 SHiELD-SOM by \qty{25}{\percent} in the 3xCO$_2$ and 4xCO$_2$ climates, and is on par with the baseline in the 2xCO$_2$ climate.

\begin{figure}
\noindent\includegraphics[width=\textwidth]{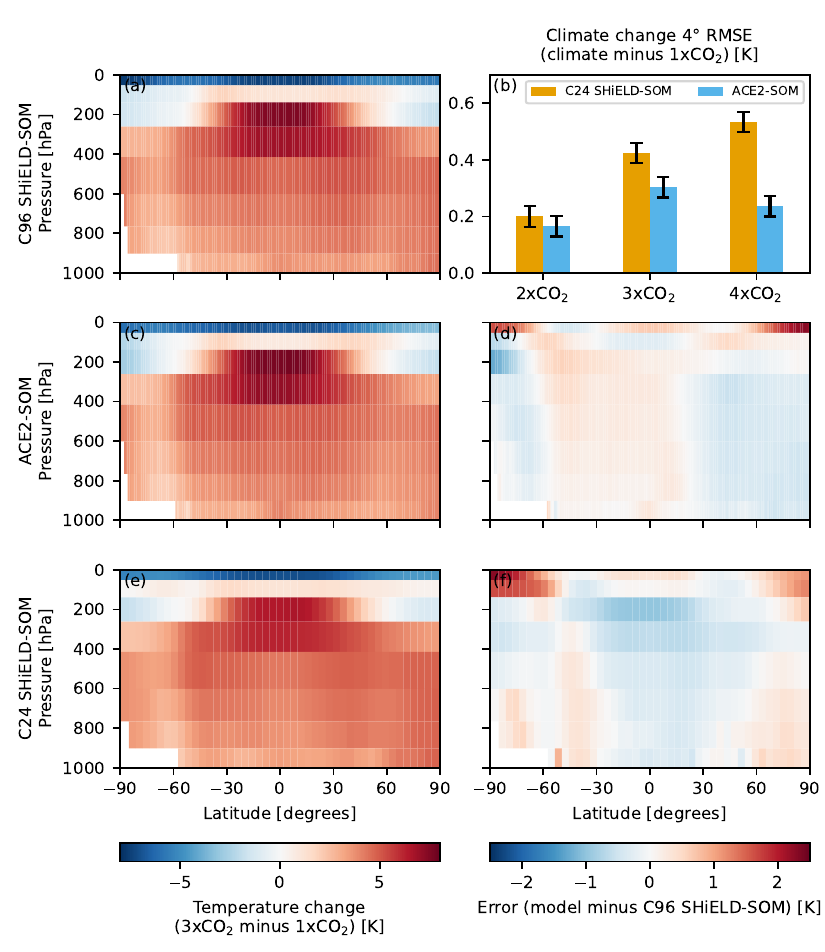}
\caption{As for Figure~\ref{fig:surface-temperature-change-3xCO2} but for changes in the zonal-mean vertical profile of temperature. The RMSE in panel (b) is computed using weights proportional to the mass of air above the surface in each zonal band and pressure level.}
\label{fig:air-temperature-change-3xCO2}
\end{figure}

If we look deeper, ACE2-SOM's skill in emulating temperature change extends from the surface to the top of the atmosphere.  Figure~\ref{fig:air-temperature-change-3xCO2} shows the vertical profile of the zonal, time, and ensemble mean temperature difference between the 3xCO$_2$ and 1xCO$_2$ climate, as well the change pattern errors exhibited by ACE2-SOM and C24 SHiELD-SOM.  Here, prior to taking a zonal mean, the temperature in each column is interpolated to a common 8 pressure levels, chosen to represent the midpoint of ACE2's vertical coordinate assuming a surface pressure of \qty{1000}{\hecto \Pa} based on Equation 3.17 of \cite{Sim1981}.  The well-known pattern of greenhouse-gas-induced warming throughout the troposphere, and cooling in the stratosphere, with warming reaching a maximum in the tropical upper troposphere, is evident in all models \cite{Man1967,Lee2021,San2023}.  ACE2-SOM's errors relative to C96 SHiELD-SOM are small nearly everywhere, the largest being too little cooling in the top layer in the polar regions, and slightly too much cooling in the high latitudes of the Southern Hemisphere in the third layer from the top.
C24 SHiELD-SOM exhibits too muted stratospheric cooling in the high latitudes of the top two layers and exhibits too muted a temperature increase in the tropical upper troposphere, resulting in a larger root mean square error (Figure~\ref{fig:air-temperature-change-3xCO2}b).

\begin{figure}
\noindent\includegraphics[width=\textwidth]{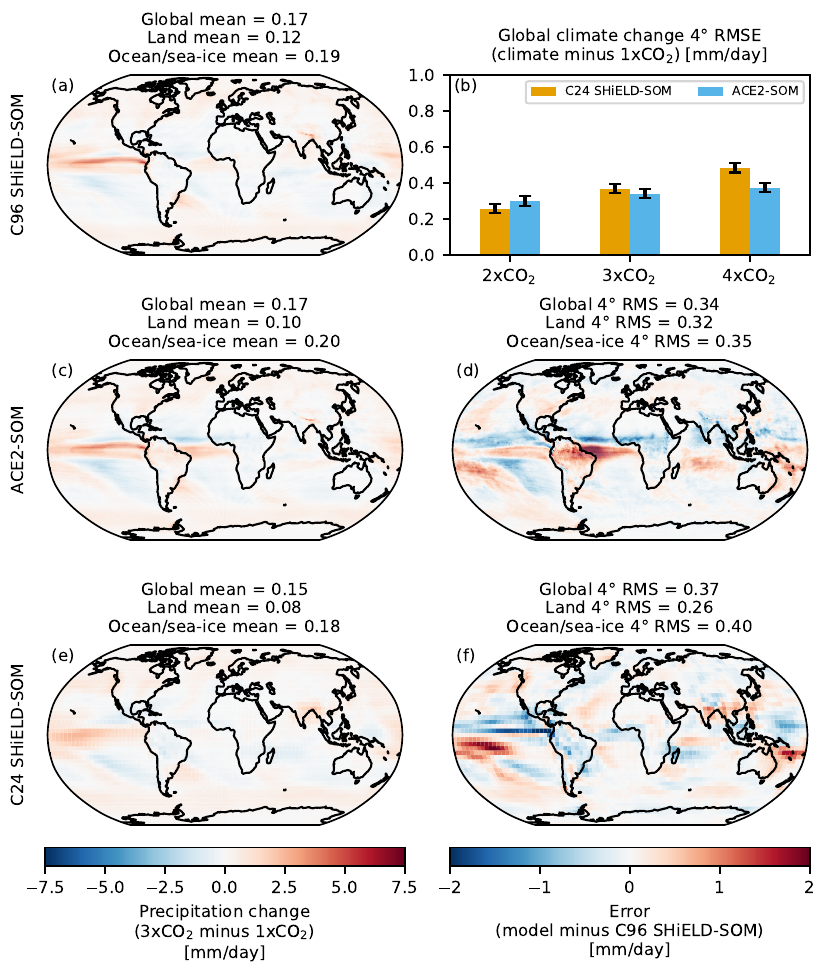}
\caption{As for Figure~\ref{fig:surface-temperature-change-3xCO2} but for precipitation.}
\label{fig:precipitation-change-3xCO2}
\end{figure}

Figure~\ref{fig:precipitation-change-3xCO2} shows the climate change patterns and pattern errors for precipitation. As in the case of surface temperature, the qualitative spatial patterns in the 3xCO$_2$ climate are similar for C96 SHiELD-SOM, ACE2-SOM, and C24 SHiELD-SOM. Changes in precipitation are modest in a global mean sense, consistent with the \qtyrange[range-units = single]{\sim 1}{3}{\percent \per \kelvin} scaling observed and physically motivated in many previous climate modeling studies \cite<e.g.,>{All2002a,Hel2006,Ste2008,Jee2018}. The most visually apparent spatial change is the ``wet-get-wetter, dry-get-drier'' pattern over ocean, with precipitation increasing within in the Intertropical Convergence Zone (ITCZ), decreasing in regions of subsidence in the subtropics, and increasing in the mid-latitude storm tracks \cite{Hel2006}.

Figures~\ref{fig:precipitation-change-3xCO2}d~and~\ref{fig:precipitation-change-3xCO2}f show the errors in simulating the climate change pattern in precipitation for the 3xCO$_2$ climate for ACE2-SOM and C24 SHiELD-SOM. Here the benefit of the 4x finer resolution of ACE2-SOM relative to C24 SHiELD-SOM is apparent in the tropical Pacific, with the increase pattern along the ITCZ being more muted and diffuse. ACE2-SOM appears to have a systematic wet bias in the Equatorial Pacific and Atlantic, and a dry bias to the north. In a global sense the change pattern RMSEs, depicted in Figure~\ref{fig:precipitation-change-3xCO2}b, are in ACE2-SOM on par with or smaller than in C24 SHiELD-SOM in the in-sample and out-of-sample equilibrium climates tested.

\begin{figure}
\noindent\includegraphics[width=\textwidth]{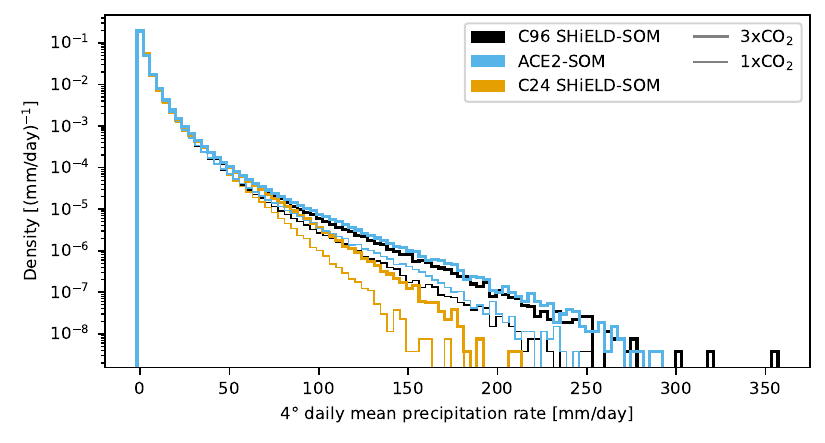}
\caption{Histograms of daily-mean precipitation rate in C96 SHiELD-SOM (black), ACE2-SOM (blue), and C24 SHiELD-SOM (orange) in the 1xCO$_2$ (thin lines) and 3xCO$_2$ (thick lines) equilibrium climates.  C96 SHiELD-SOM and ACE2-SOM data has been regridded to \qty{4}{\degree} resolution for a fair comparison with C24 SHiELD-SOM.}
\label{fig:precipitation-pdfs}
\end{figure}

While mean precipitation increases modestly with warming, extreme precipitation increases more rapidly.  We can get a sense for this by looking at Figure~\ref{fig:precipitation-pdfs}, which shows histograms of the daily mean precipitation rate with data regridded to a \qty{4}{\degree} Gaussian grid for each model in the 1xCO$_2$ and 3xCO$_2$ climates.  The tails of the distributions in each of the models, corresponding to high quantiles, increase by roughly \qty{20}{\percent}, or about \qty{6}{\percent \per \K} global mean warming, consistent with the general picture of prior studies \cite<e.g.,>{OG2009}.  At this horizontal scale, ACE2-SOM emulates C96 SHiELD-SOM fairly well across the distributions in each climate. At \qty{1}{\degree} resolution ACE2-SOM more noticeably underestimates the frequency of the most extreme precipitation events with intensities in the top millionth of the distribution, when compared with C96 SHiELD-SOM (Figure~S1).  C24 SHiELD-SOM's low precipitation bias is evident, with precipitation rates failing to reach even what they are in the 1xCO$_2$ climate of C96 SHiELD-SOM or ACE2-SOM in its 3xCO$_2$ climate, though it exhibits roughly the expected scaling behavior with warming.  Overall this suggests that ACE2-SOM is not only learning to emulate the mean precipitation change with warming, but also learning to emulate how its distribution will change.

\subsection{Non-equilibrium climate inference}

We have shown that ACE2-SOM is skilled at emulating mean equilibrium climate with CO$_2$ concentrations between 1xCO$_2$ and 4xCO$_2$. We now transition to more challenging out-of-sample test cases, where the atmospheric and oceanic state is not in equilibrium with the CO$_2$ concentration.

\subsubsection{Gradually increasing CO$_2$}

\begin{figure}
\noindent\includegraphics[width=\textwidth]{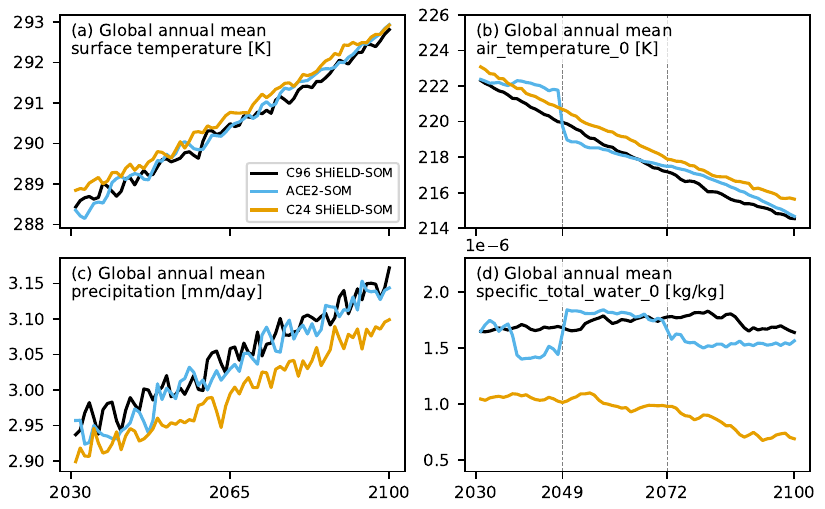}
\caption{Time evolution of global annual mean surface temperature (a), stratospheric temperature (b), precipitation rate (c), and stratospheric specific total water (d) in C96 SHiELD-SOM (black), ACE2-SOM (blue), and C24 SHiELD-SOM (orange). The vertical dashed lines at years 2049 and 2072 in panels (b) and (d) highlight that the regime shifts in stratospheric temperature and specific total water are correlated in time.}
\label{fig:gradual-co2-increase-case}
\end{figure}

In our first non-equilibrium-climate test case, CO$_2$ increases at a rate of \qty{2}{\percent \per \year} for \qty{70}{\years}. This case is analogous to the traditional CMIP experiment where CO$_2$ is prescribed to increase at a rate of \qty{1}{\percent \per \year} \cite{Eyr2016}, but reaches 4x present-day CO$_2$ at a faster rate to reduce the amount of compute time needed to run the reference SHiELD-SOM simulations. Figure~\ref{fig:gradual-co2-increase-case} shows the time evolution of the global and annual mean of four fields in simulations with C96 SHiELD-SOM, ACE2-SOM, and C24 SHiELD-SOM. Panels~\ref{fig:gradual-co2-increase-case}a~and~\ref{fig:gradual-co2-increase-case}c depict surface temperature and precipitation rate, which are examples of fields that ACE2-SOM emulates well in this context. Generally the global annual mean curves of ACE2-SOM follow the trend of C96 SHiELD-SOM, with reasonable interannual variability; the systematic bias of the baseline C24 SHiELD-SOM simulation is evident, particularly for precipitation.  Other variables with meaningful global means generally also look reasonable with ACE2-SOM (not shown).

However, panels~\ref{fig:gradual-co2-increase-case}b~and~\ref{fig:gradual-co2-increase-case}d depict stratospheric air temperature and specific total water, which are the fields that ACE2-SOM emulates least well. Consistent with increased stratospheric radiative cooling as CO$_2$ increases \cite{San2023}, stratospheric air temperature decreases at a steady rate in C96 and C24 SHiELD-SOM. In ACE2-SOM, however, it decreases at a muted rate, then decreases abruptly in year 2049, then decreases with a muted rate again until it increases slightly in 2072, and finally decreases with a slightly accelerated rate for the remainder of the run. Stratospheric specific total water has little discernible trend in our target C96 SHiELD-SOM, with a global annual mean meandering between \qtyrange[range-units = single]{1.64e-6}{1.83e-6}{\kg \per \kg} throughout the run. ACE2-SOM roughly captures this qualitative behavior—the stratospheric specific total water at the end of the run is similar to what it was at the beginning—but exhibits large regime shifts around the same time as the stratospheric temperature. C24 SHiELD-SOM exhibits a large dry bias and overall drying trend, decreasing by roughly \qty{30}{\percent} by the end of the simulation.  

We speculate these regime shifts are a result of correlations between the quantized CO$_2$ concentrations in our equilibrium climate training data and these slowly varying stratospheric variables. In other words ACE2-SOM learns to associate certain ranges of CO$_2$ with certain values of stratospheric specific total water, and to a lesser extent stratospheric air temperature. For example, it happens that the stratospheric specific total water in the equilibrium 2xCO$_2$ climate training data is larger than it is in the 1xCO$_2$ and 4xCO$_2$ equilibrium climates (\qty{1.78e-6}{\kg \per \kg} versus \qty{1.66e-6}{\kg \per \kg} and \qty{1.62e-6}{\kg \per \kg}, respectively), which is qualitatively consistent with how ACE2-SOM predicts it will evolve as CO$_2$ varies between 1xCO$_2$ and 4xCO$_2$. Global-mean stratospheric total water and air temperature are accurate halfway through the simulation, where the in-sample 2xCO2 value is used. The regime shifts may occur because ACE2-SOM is learning to predict mainly the climatology of stratospheric specific total water based on the CO$_2$ concentration, rather than how it will evolve over a six-hour time interval.  This is consistent with the fact that regime shifts in increasing CO$_2$ inference runs are less common or extreme with models trained on output from the increasing CO$_2$ simulation, but notably they are not entirely absent (Section~\ref{sec:ramped-training}).

\subsubsection{Abrupt CO$_2$ increase}

Another CMIP DECK experiment consists of an abrupt quadrupling of CO$_2$ from an equilibrium climate \cite{Eyr2016}. Here we describe the results of attempting a similar experiment with ACE2-SOM. This kind of simulation is normally run for at least \qty{150}{\years} in fully coupled models, as it takes the deep ocean many years to equilibrate \cite<see the motivation for>{Rug2019}. A slab ocean, on the other hand, equilibrates more quickly, so it is sufficient to look at 10-year runs in our case. 

\begin{figure}
\noindent\includegraphics[width=\textwidth]{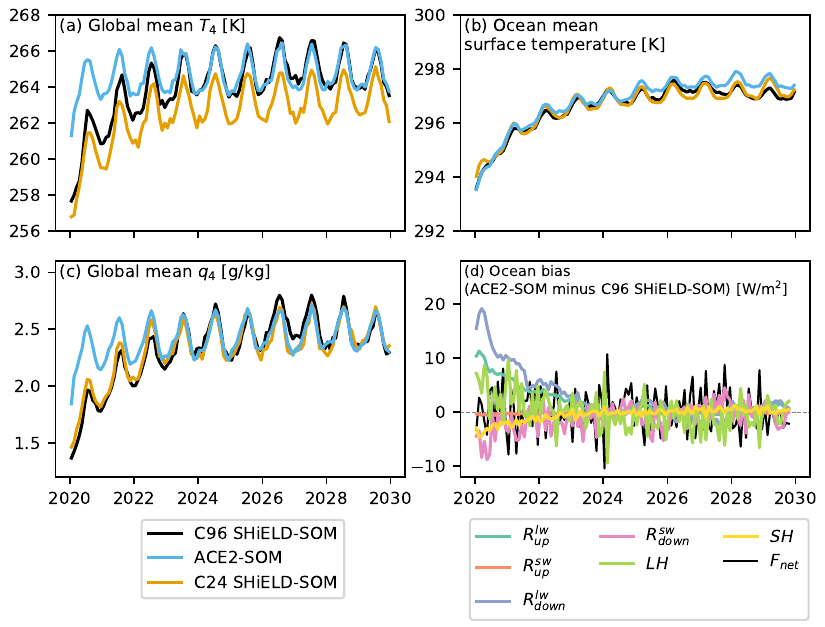}
\caption{Time evolution of global monthly mean temperature at the fourth vertical level, numbered from top of atmosphere to bottom (a), ocean monthly mean surface temperature (b), and global monthly mean specific total water at the fourth vertical level (c) in a simulations where the CO$_2$ concentration was abruptly changed from 1xCO$_2$ to 4xCO$_2$ at the start of the run. Panel (d) shows the time evolution of the bias (ACE2-SOM minus C96 SHiELD-SOM) in the ocean monthly mean of the components of $F_{net}$, the atmosphere's coupling mechanism with the slab ocean, as well as $F_{net}$ itself.  For simplicity of interpretation, ``ocean mean'' pertains to the mean taken over grid cells that are 100\% ocean throughout the year.}
\label{fig:abrupt-4xco2}
\end{figure}

This is a challenging out-of-sample test for ACE2-SOM, due to its highly non-equilibrium character.  When CO$_2$ is abruptly quadrupled, the slab ocean response is reasonable (Figure~\ref{fig:abrupt-4xco2}b).  However, all directly ML-predicted atmospheric fields rapidly shift to what their values would be in a 4xCO$_2$ equilibrium climate simulation, as seen in time series of the global and monthly mean mid-tropospheric temperature (Figure~\ref{fig:abrupt-4xco2}a) and specific total water (Figure~\ref{fig:abrupt-4xco2}c), which deviate substantially from the trajectories of those in the physics-based C96 and C24 SHiELD-SOM simulations in their first \qty{3}{\years}. 

\begin{figure}
\noindent\includegraphics[width=\textwidth]{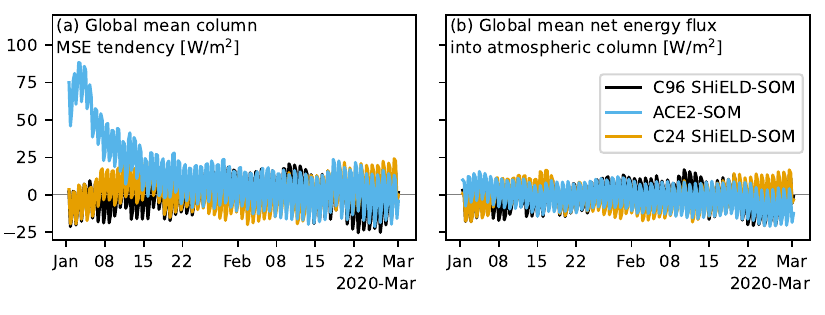}
\caption{6-hourly global mean column-integrated moist static energy tendency (a) and net energy flux into the atmosphere (b) in the first two months of an inference run with abruptly quadrupled CO$_2$ with C96 SHiELD-SOM (black), ACE2-SOM (blue), and C24 SHiELD-SOM (orange).}
\label{fig:abrupt-4xco2-energy-non-conservation}
\end{figure}

The abrupt regime shift of these variables in ACE2-SOM in this experiment is not physically realistic, because it violates global energy conservation. Figure~\ref{fig:abrupt-4xco2-energy-non-conservation} illustrates this by plotting the time series of the global mean column integrated moist static energy tendency side by side with the net column energy input into the atmosphere in the first two months of the simulation. In an approximately energy conserving model the two curves would line up, according to the moist static energy budget:
\begin{equation}
    \frac{ \partial \left\{ \left< m \right> \right\} }{\partial t} = \left\{ F^{toa}_{net} \right\} - \left\{ F^{sfc}_{net} \right\}
\end{equation}
where $m$ is the moist static energy, the angle brackets indicate a mass-weighted vertical integral, the curly braces denote a global area-weighted mean, and $F^{toa}_{net}$ and $F^{sfc}_{net}$ are the net downward energy fluxes at the top of the atmosphere and surface \cite{Nee1987}. The curves in Figure~\ref{fig:abrupt-4xco2-energy-non-conservation}a~and~Figure~\ref{fig:abrupt-4xco2-energy-non-conservation}b approximately do line up in the case of C96 and C24 SHiELD-SOM, but clearly do not in the case of ACE2-SOM. In ACE2-SOM there is a rapid heating and moistening of the atmosphere that is not supported by an equivalent net energy input through its boundaries. This is in line with the hypothesis that while exhibiting a small amount of thermal inertia, the model is mainly attempting to predict an accurate climatology given the CO$_2$ concentration, rather than an accurate time evolution, during these out-of-sample forcing periods.

Even the realistic rate of warming of the slab ocean in ACE2-SOM is not occurring for the right reason.
While the sign and magnitude of the predicted $F_{net}$ is roughly consistent with that of C96 SHiELD-SOM, it is a result of largely compensating biases in its components, shown in Figure~\ref{fig:abrupt-4xco2}d. The most biased among these are the downward and upward longwave radiative fluxes, which are both biased high, partially offsetting, as well as the downward shortwave radiative flux and latent heat flux, which also both act to offset the positive bias in the downward longwave radiative flux. One could argue that the positive bias in downward longwave radiative flux is at least qualitatively consistent with the positive bias in the temperature of the atmosphere; however, the positive bias in upward longwave radiative flux is not physically consistent with the only slightly warmer ocean.  Based on a linearization of the Stefan-Boltzmann Law about \qty{294}{\K}, a bias in upward longwave radiative flux at the surface of \qtyrange[range-units = single]{5}{10}{\W \per \meter \squared} would require a temperature bias roughly between \qtyrange[range-units = single]{0.87}{1.74}{\K}, which is greater than that exhibited at any point throughout the run. This suggests that ACE2-SOM has spuriously learned that the upward longwave radiative flux at the surface depends not only on the surface temperature, but also the concentration of CO$_2$ and other properties of the atmosphere, since these co-vary in the training dataset.

\subsubsection{Radiation multi-call experiments}

\begin{figure}
 \noindent\includegraphics[width=\textwidth]{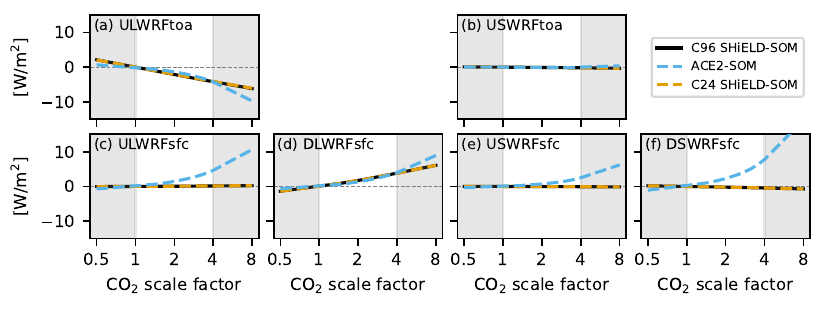}
\caption{One-year mean difference between radiative flux components predicted with CO$_2$ perturbed by a varying scale factor and those predicted with 1xCO$_2$. Regions with gray shading correspond to CO$_2$ concentrations that are outside the range seen during training.}
\label{fig:radiation-multi-calls}
\end{figure}

Learning this unphysical relationship between the upward longwave radiative flux at the surface and other fields is likely a result of ACE2-SOM's lack of exposure to non-equilibrium combinations of CO$_2$ concentrations and atmospheric states during training. We can illustrate this issue more directly through radiation multi-call experiments typically used for computing ``instantaneous radiative forcings'' \cite<e.g.,>{Pin2020}, which we can perform with both SHiELD and ACE2-SOM. In these experiments, the top of atmosphere and surface radiative fluxes are predicted with identical atmospheric states, but varying CO$_2$ concentrations. Figure~\ref{fig:radiation-multi-calls} shows the difference in one-year mean radiative fluxes for CO$_2$ scaled by a varying factor and the control 1xCO$_2$ in C96 SHiELD-SOM, ACE2-SOM, and C24 SHiELD-SOM. The upward longwave radiative flux at the top of the atmosphere and the downward longwave radiative flux at the surface are the only variables which should have a meaningful physical response to changing CO$_2$, scaling approximately with the logarithm of the concentration \cite<cf. Figure 1 of>{Hua2014a}.  C96 SHiELD-SOM and C24 SHiELD-SOM exhibit this well (Figures~\ref{fig:radiation-multi-calls}a~and~\ref{fig:radiation-multi-calls}d). Consistent with the greenhouse effect, as the CO$_2$ increases, the upward longwave radiative flux at the top of the atmosphere decreases and the downward longwave radiative flux at the surface increases. ACE2-SOM approximately emulates this, even with CO$_2$ concentrations outside the range seen during training, albeit missing the logarithmic dependence on CO$_2$. On the other hand, Figures~\ref{fig:radiation-multi-calls}c,~\ref{fig:radiation-multi-calls}e,~and~\ref{fig:radiation-multi-calls}f all correspond to fields that should not physically depend on the CO$_2$ concentration—shortwave radiative fluxes, as well as the upward longwave radiative flux at the surface—but ACE2-SOM predicts that they do. While this seed of ACE2-SOM predicts little sensitivity of the upward shortwave radiative flux at the top of the atmosphere to CO$_2$ (Figure~\ref{fig:radiation-multi-calls}b), which is physically realistic, this appears to be due to chance, as other seeds exhibit a less trivial sensitivity (not shown). 

\subsection{Training on the collection of equilibrium climate runs versus on the increasing-CO$_2$ run}
\label{sec:ramped-training}

A potential alternative training strategy would be to train on output from the increasing CO$_2$ run instead of output from the collection of equilibrium climate runs.  This would expose ACE2-SOM to a less quantized range of climate states and CO$_2$ concentrations, a diversity which could potentially be beneficial, though these states would not quite be in equilibrium. Here we investigate the sensitivity of ACE2-SOM's skill in equilibrium-climate and increasing-CO$_2$ inference to its training and checkpoint selection strategy; this also gives us an opportunity to discuss random seed variability.  For this purpose, we train four models on output from the increasing-CO$_2$ run, holding out the middle \qty{14}{\years} for validation and out-of-sample testing, corresponding to CO$_2$ concentrations between 1.74xCO$_2$ and 2.25xCO$_2$.  We choose our best checkpoint during training based on results of eight \qty{7}{\year} inference simulations with initial conditions selected to evenly cover all \qty{56}{\years} of increasing-CO$_2$ training data.

\begin{figure}
\noindent\includegraphics[width=\textwidth]{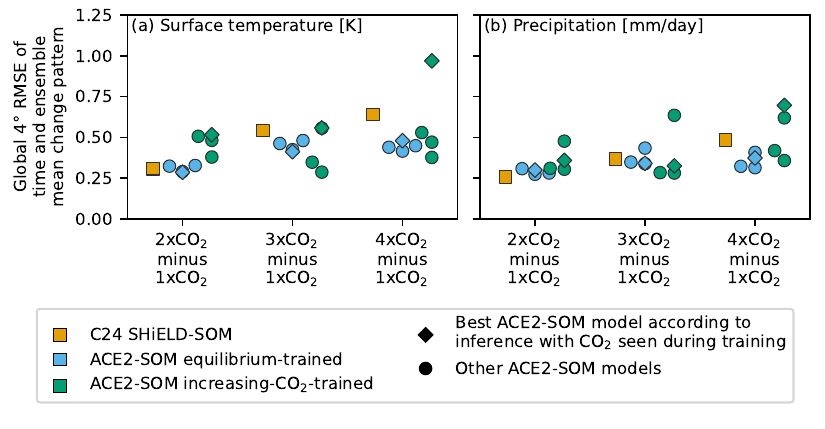}
\caption{Global \qty{4}{\degree} root mean square error of the time and ensemble mean climate change pattern of surface temperature (a) and precipitation (b) with C24 SHiELD-SOM (orange), ACE2-SOM trained only on equilibrium climate data (blue), and ACE2-SOM trained only on output from the increasing-CO$_2$ run (green). Diamonds represent results with ACE2-SOM models chosen based on performance in inference with CO$_2$ concentrations seen during training; circles represent results with ACE2-SOM models trained with different random seeds.}
\label{fig:rs-variability-eq-climate}
\end{figure}

Figure~\ref{fig:rs-variability-eq-climate} provides a high-level overview of the skill of ACE2-SOM in emulating the equilibrium climate change patterns of surface temperature and precipitation with these different training approaches. The sample size is small—four random seeds per training approach—but the equilibrium-climate-trained models appear to improve upon the C24 SHiELD-SOM baseline slightly more consistently than the increasing-CO$_2$-trained models. Perhaps as a result of the checkpoint selection strategy based on inference with CO$_2$ forcings seen during training, there is also a greater spread in skill across seeds in the equilibrium-climate context for the increasing-CO$_2$-trained models.  In particular, the best model/checkpoint chosen based on inline inference in the increasing-CO$_2$ climate happens to be one of the poorer performing models for these metrics.

\begin{figure}
\noindent\includegraphics[width=\textwidth]{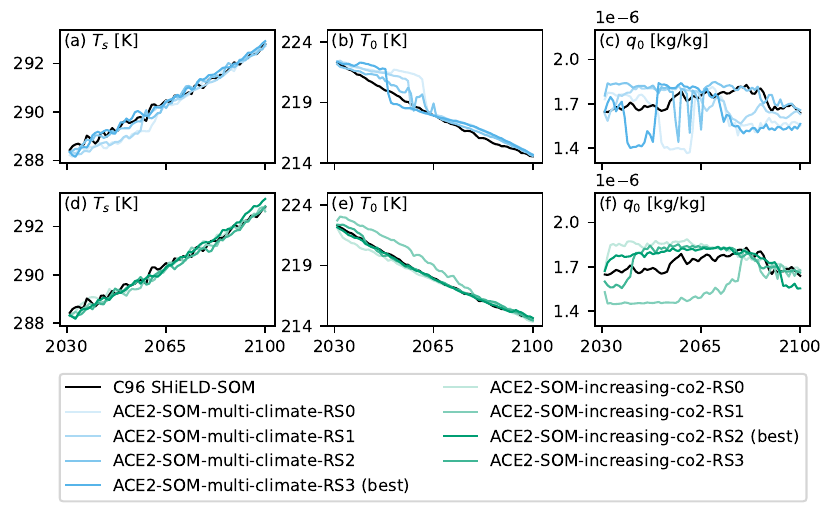}
\caption{Global annual mean time series of surface temperature (a) and (d), stratospheric temperature (b) and (e), and stratospheric specific total water (c) and (f) in equilibrium-climate-trained models (top row) and increasing-CO$_2$-trained models (bottom row).  The target C96 SHiELD-SOM results are shown in black in each panel.  The model chosen by checkpoint selection is the darkest and most-foregrounded line; lighter shaded lines correspond to other seeds.}
\label{fig:rs-variability-increasing-co2}
\end{figure}

If we look at skill in increasing-CO$_2$ inference, shown in Figure~\ref{fig:rs-variability-increasing-co2}, we find unsurprisingly that increasing-CO$_2$-trained models tend to slightly outperform equilibrium-climate-trained models. All increasing-CO$_2$-trained models produce a smooth time series of global annual mean surface temperature (Figure~\ref{fig:rs-variability-increasing-co2}d), in comparison to two equilibrium-climate-trained models, which exhibit a negative bias in the first half of the run (Figure~\ref{fig:rs-variability-increasing-co2}a).  For stratospheric temperature and specific total water, however, the increasing-CO$_2$-trained models can still suffer from similar pitfalls as the equilibrium-climate-trained models.  While $q_0$ varies somewhat more smoothly in Figure~\ref{fig:rs-variability-increasing-co2}f than in Figure~\ref{fig:rs-variability-increasing-co2}c, some models still produce regime-shift behavior that maps onto periods of bias in $T_0$ (Figure~\ref{fig:rs-variability-increasing-co2}e).

Importantly, however, neither of these training strategies provides qualitatively different behavior in the abrupt 4xCO$_2$ scenario, or sensitivities exhibited in the radiation multi-call experiments (not shown).  This suggests that changes to the training dataset and/or formulation of ACE2-SOM will be needed to achieve success in such tests.

\section{Discussion and Conclusion}
\label{sec:discussion-and-conclusion}

In this study we have shown that ACE2 coupled to a slab ocean model can be trained to successfully emulate the equilibrium climate of a physics-based climate model with varying CO$_2$ concentration.  Like earlier versions of ACE—which were aided by prescribed sea surface temperatures—ACE2-SOM is highly stable with annually repeating forcings, exhibiting realistic interannual variability in rollouts. In individual climates, ACE2-SOM strongly outperforms a 4x coarser, yet 25 times more energy intensive, physics-based baseline model in emulating the time-mean pattern of the target \qty{100}{\km} resolution model.  In emulating climate change patterns, for which biases of the baseline model largely cancel out, ACE2-SOM outperforms or is at least on par with the baseline. This is a remarkable pilot demonstration of the potential of a machine learning emulator of a climate model for accurate, computationally efficient simulation of anthropogenic climate perturbations.  To be fully competitive with physics-based climate models, however, ACE2-SOM's ability to emulate out-of-sample conditions, such as non-equilibrium climates, needs future improvements in model formulation and likely in choice of training data.  This provides many interesting directions for ongoing research.

Additionally there is a need for developing an emulator that realistically includes important additional components of the Earth system, such as the circulation of the ocean and coverage of sea ice, both of which can amplify the equilibrium climate sensitivity of surface temperature to changes in CO$_2$ \cite<e.g.,>{Dun2020,Hal2004}.  However, by analogy with the development of physics-based models, there is still much we can learn about emulating the response of climate to changes in the composition of the atmosphere even with a slab ocean approach like the one used here.  Beyond the specific questions related to the test cases in this study, some broader open questions are, how might we move beyond emulating the forced response to a single well-mixed greenhouse gas?  How might training approaches need to differ to capture the response to spatially heterogeneous emissions and resulting atmospheric burdens of aerosols?  How might we handle emulating the response to combinations of forcings?  Answering these questions in a parsimonious and physically interpretable way will help machine-learning based emulators become credible tools for projecting climate change under different emissions scenarios, and is something that can be pursued in parallel to extending emulation to include other components of the Earth system.

\section*{Open Research Section}

The code used for data processing, model training, inference, and evaluation is available at \url{https://github.com/ai2cm/ace} \cite{Wat2024a}. The scripts used for submitting experiments and generating figures are available at \url{https://github.com/ai2cm/ace2-som-paper} \cite{Cla2024}. Processed reference data from SHiELD-SOM used for training and testing ACE2-SOM can be found in the following public requester-pays bucket in Google Cloud Storage: \url{gs://ai2cm-public-requester-pays/2024-12-05-ai2-climate-emulator-v2-som}.  Finally, the checkpoint of the best equilibrium-climate-trained model discussed in this manuscript can be found on Hugging Face along with sample reference forcing data at \url{https://huggingface.co/allenai/ACE2-SOM}.

\acknowledgments

We acknowledge NOAA's Geophysical Fluid Dynamics Laboratory for the computing resources used to complete the SHiELD-SOM reference simulations.  This research also used resources of NERSC, a U.S. Department of Energy Office of Science User Facility located at Lawrence Berkeley National Laboratory, using NERSC award BER-ERCAP0026743.  We thank Kun Gao, Baoqiang Xiang, and Linjiong Zhou for discussion and review of code modifications needed for running with a global slab ocean in SHiELD, and Wenhao Dong for reviewing an earlier version of this manuscript. Finally, we thank NOAA's Environmental Modeling Center for making GFS analysis, reference data, and the necessary software available for producing initial conditions and forcing data for SHiELD at different resolutions.

%
%

\bibliography{references.bib}

%
%
%
%
%

\end{document}


%
%

\title{Supporting Information for ``ACE2-SOM: Coupling an ML atmospheric emulator to a slab ocean and learning the sensitivity of climate to changed CO$_2$''}

%
%

\authors{Spencer K. Clark\affil{1,2}, Oliver Watt-Meyer\affil{1}, Anna Kwa\affil{1}, Jeremy McGibbon\affil{1}, Brian Henn\affil{1}, W. Andre Perkins\affil{1}, Elynn Wu\affil{1}, Lucas M. Harris\affil{2}, and Christopher S. Bretherton\affil{1}}


\affiliation{1}{Allen Institute for Artificial Intelligence, Seattle, WA, USA}
\affiliation{2}{NOAA/Geophysical Fluid Dynamics Laboratory, Princeton, NJ, USA}


%
%

%

\begin{article}

%
%

\noindent\textbf{Contents of this file}
\begin{enumerate}
\item Table S1
\item Figure S1
\end{enumerate}

%
%
\end{article}


\begin{table}
  \caption{Input and output variables for ACE2-SOM. Table, caption, and notation are adapted from \cite{Wat2024}. The $k$ subscript refers to a vertical layer index, and ranges from 0 to 7 starting at the top of atmosphere and increasing towards the surface. The Time column indicates whether a variable represents the value at a particular time step (``Snapshot''), the average across the 6-hour time step (``Mean'') or a quantity which does not depend on time (``Invariant''). ``TOA'' denotes ``Top Of Atmosphere'', the climate model's upper boundary.}
  \label{table:variables}
  \centering
  \begin{tabular}{llll}
    \\
    \multicolumn{4}{c}{Prognostic (input and output)}   \\
    \hline
    Symbol   & Description                                & Units & Time        \\
    \hline
    $T_k$    & Air temperature                            & K     & Snapshot \\
    $q^T_k$   & Specific total water (vapor + condensates) & kg/kg & Snapshot \\
    $u_k$    & Windspeed in eastward direction            & m/s   & Snapshot \\
    $v_k$    & Windspeed in northward direction           & m/s   & Snapshot \\
    $T_s$     & Skin temperature           & K     & Snapshot \\
    $p_s$     & Atmospheric pressure at surface            & Pa    & Snapshot \\
    $T_{2m}$  & 2-meter air temperature                     & K    & Snapshot \\
    $q_{2m}$  & 2-meter specific humidity                     & kg/kg    & Snapshot \\
    $u_{10m}$  & 10-meter windspeed in eastward direction      & m/s    & Snapshot \\
    $v_{10m}$  &  10-meter windspeed in northward direction      & m/s    & Snapshot \\
    \hline
    \\
    \multicolumn{4}{c}{Forcing (input only)}   \\
    \hline
    Symbol       & Description                              & Units   & Time          \\
    \hline
    DSWRF$_{toa}$ & Downward shortwave radiative flux at TOA & W/m$^2$ & Mean  \\
    $z_s$       & Surface height of topography             & m      & Invariant \\
    $f_l$       & Land grid cell fraction                  & $-$    & Invariant \\
    $f_o$       & Ocean grid cell fraction                 & $-$    & Snapshot \\
    $f_{si}$    & Sea-ice grid cell fraction               & $-$    & Snapshot \\
    $\mathrm{CO}_2$ & Global mean atmospheric carbon dioxide & ppm  & Snapshot \\
    \hline
    \\
    \multicolumn{4}{c}{Diagnostic (output only)}   \\
    \hline
    Symbol       & Description                                    & Units   & Time    \\
    \hline
    USWRF$_{toa}$ & Upward shortwave radiative flux at TOA         & W/m$^2$ & Mean \\
    ULWRF$_{toa}$ & Upward longwave radiative flux at TOA          & W/m$^2$ & Mean  \\
    USWRF$_{sfc}$ & Upward shortwave radiative flux at surface     & W/m$^2$ & Mean  \\
    ULWRF$_{sfc}$ & Upward longwave radiative flux at surface      & W/m$^2$ & Mean  \\
    DSWRF$_{sfc}$ & Downward shortwave radiative flux at surface   & W/m$^2$ & Mean  \\
    DLWRF$_{sfc}$ & Downward longwave radiative flux at surface    & W/m$^2$ & Mean  \\
    $P$ & Surface precipitation rate (all phases)               & kg/m$^2$/s & Mean  \\
    $\left. \frac{\partial TWP}{\partial t}\right |_{adv}$ & Tendency of total water path from advection & kg/m$^2$/s & Mean \\
    $LHF$ & Surface latent heat flux                            & W/m$^2$ & Mean  \\
    $SHF$ & Surface sensible heat flux                          & W/m$^2$ & Mean  \\
    $Z_{500}$ & 500$\,$hPa geopotential height                    & m & Snapshot   \\
    $T_{850}$ & 850$\,$hPa air temperature                       & K & Snapshot   \\
    \hline
  \end{tabular}
\end{table}

\begin{figure}
\noindent\includegraphics{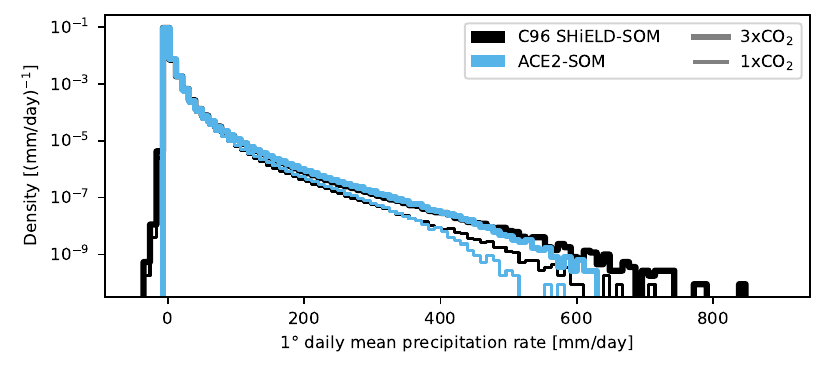}
\caption{Histograms of daily-mean precipitation rate in C96 SHiELD-SOM (black) and ACE2-SOM (blue) in the 1xCO$_2$ (thin lines) and 3xCO$_2$ (thick lines) equilibrium climates at \qty{1}{\degree} resolution.}
\label{fig:one-degree-precipitation-histograms}
\end{figure}
\clearpage
\bibliography{references.bib}